\documentclass[journal]{IEEEtran}

  \usepackage{graphicx}
    \usepackage{amsmath}
    \usepackage{amssymb}
    \usepackage{array}
    \usepackage{multicol}
    \usepackage{bm}
    \usepackage{multirow}
    \usepackage{cite}
    \usepackage{color}

\newcommand{\LB}{\left(}
\newcommand{\RB}{\right)}
\newcommand{\LSB}{\left[}
\newcommand{\RSB}{\right]}

\newcommand{\lv}{\lvert}
\newcommand{\rv}{\rvert}

\newcommand{\htp}{^{\sf H}}

\newcommand{\CC}{{\mathbb C}}
\newcommand{\RR}{{\mathbb R}}
\newcommand{\EE}{{\mathbb E}}
\newcommand{\PP}{{\mathbb P}}

\newcommand{\g}{{\bf g}}
\newcommand{\h}{{\bf h}}

\newcommand{\m}{{\bf m}}

\newcommand{\w}{{\bf w}}
\newcommand{\x}{{\bf x}}

\newcommand{\zerov}{{\bf 0}}

\newcommand{\G}{{\bf G}}

\newcommand{\I}{{\bf I}}

\newcommand{\Cc}{{\cal C}}

\newcommand{\Lc}{{\cal L}}

\newcommand{\Nc}{{\cal N}}
\newcommand{\Oc}{{\cal O}}

\newcommand{\Sc}{{\cal S}}

\newtheorem{rem}{Remark}

\newcommand{\mr}{\mathrm}

\newtheorem{theorem}{Theorem}

\begin{document}
\title{
Optimizing the MIMO Cellular Downlink: 
Multiplexing, Diversity, or Interference Nulling?}
\author{
Kianoush Hosseini, Caiyi Zhu, Ahmad Khan, Raviraj S. Adve, and Wei Yu%
\thanks{This work was supported in part by the Natural Sciences and Engineering Research Council (NSERC) via the CRD, Discovery Grant and Canada Research Chair programs and in part by TELUS Canada. This paper has been presented in part at the Allerton Conf.~on Comm., Control and Computing, 2015~\cite{HYA_Allerton15}. 

K.~Hosseini and C.~Zhu were with The Edward S.~Rogers Sr.\ Department of Electrical and Computer Engineering, University of Toronto, Toronto, ON M5S 3G4, Canada. K.~Hosseini is now with Qualcomm Research Center, San Diego, CA, 92121 (email: kianoush.hosseini@alum.utoronto.ca). C.~Zhu is now with Google, Waterloo, ON (email: caiyi.zhu@mail.utoronto.ca). A. Khan, R.~S.~Adve and W.~Yu are with The Edward S.~Rogers Sr.\ Department of Electrical and Computer Engineering, University of Toronto, Toronto, ON M5S 3G4, Canada (emails: ahmadali.khan@mail.utoronto.ca, \{rsadve, weiyu\}@comm.utoronto.ca).
}
}

\markboth{}
{Hosseini, et al: Optimizing the Large-Scale MIMO Downlink}

\maketitle
\vspace{-0.3in}
\begin{abstract}
A base-station (BS) equipped with multiple antennas can use its spatial dimensions in three different ways: (1) to serve multiple users, thereby achieving a multiplexing gain, (2) to provide spatial diversity in order to improve user rates and (3) to null interference in neighboring cells. This paper answers the following question: What is the optimal balance between these three competing benefits? We answer this question in the context of the downlink of a cellular network, where multi-antenna BSs serve multiple single-antenna users using zero-forcing beamforming with equal power assignment, while nulling interference at a subset of out-of-cell users. Any remaining spatial dimensions provide transmit diversity for the scheduled users. Utilizing tools from stochastic geometry, we show that, surprisingly, to maximize the per-BS ergodic sum rate, with an optimal allocation of spatial resources, interference nulling does not provide a tangible benefit. The strategy of \emph{avoiding} inter-cell interference nulling,  reserving some fraction of spatial resources for multiplexing and using the rest to provide diversity, is already close-to-optimal in terms of the sum-rate. However, interference nulling does bring significant benefit to cell-edge users, particularly when adopting a range-adaptive nulling strategy where the size of the cooperating BS cluster is increased for cell-edge users.
\end{abstract}

\section{Introduction}

Multiple-input multiple-output (MIMO) basestations (BS) provide myriad possibilities in optimizing transmissions in wireless cellular networks. Indeed, the recent focus of MIMO systems has been on BSs with a large, even an asymptotically large, number of antennas (massive MIMO systems). The availability of multiple antennas provides the flexibility to achieve multiple system objectives simultaneously: (1) to serve multiple users in the same time-frequency slot to achieve a spatial multiplexing gain, (2) to provide spatial diversity for these users, and, as often stated as a major advantage of massive MIMO systems, (3) to null interference at out-of-cell users. In fact, as often pointed out in the massive MIMO literature, in the limit as the number of BS antennas goes to infinity, transmit beamformers matched to the channel completely eliminate multiuser and intercell interference, leaving pilot contamination as the only limiting factor~\cite{M10}.

This paper focuses on the design of large-scale MIMO (LS-MIMO)
downlink systems~\cite{HYA14}, where each BS is equipped with a large but 
{\it finite} number of antennas\footnote{While we do not make any assumptions as to the number of BS antennas, our question is most relevant to a system with a sufficiently large number of antennas such that the three uses, multiplexing, diversity and interference nulling, are possible simultaneously.}. Specifically, we consider a scenario in which the number of users being served is comparable to or is a significant fraction of the number of BS antennas; in this case, the overall system does not necessarily operate in the so-called massive MIMO regime. With a finite number of antennas and substantial number of scheduled users, it is crucial to understand at a system level the trade-offs between multiplexing, diversity and interference cancellation. To further this understanding, we ask the following key question: between providing spatial multiplexing, diversity and mitigating intercell interference, which takes priority? Equivalently, what is the optimal trade-off among the three?

To answer this question, we use tools from stochastic geometry to analyze LS-MIMO systems assuming that BSs are spatially distributed according to a homogeneous Poisson point process (PPP). We adopt a user-centric clustering strategy in which each user receives its desired signal from the closest BS, while requesting interference nulling from a cluster of nearby BSs. To focus on the question at hand, we assume that perfect channel state information (CSI) is available without incurring any cost, at both the BSs and the users. Each BS serves multiple, single-antenna, users simultaneously using zero-forcing (ZF) beamforming with equal power allocation across the beams. Users are scheduled in a round-robin fasion. The remaining spatial dimensions at each BS are used to provide diversity for the scheduled users or to null interference at a subset of out-of-cell users.

For a single-tier wireless cellular network, we arrive at the following surprising conclusion: In terms of maximizing the per-BS ergodic sum rate, devoting spatial dimensions to interference nulling \emph{does not provide a significant benefit}. In fact, it is near-optimum, in sum-rate sense, to use a fraction of the available spatial dimensions for multiplexing (i.e., with $M$ antennas at each BS, serve $K < M$ users) and to use the remaining dimensions for spatial diversity, thereby leaving \emph{none} of the spatial dimensions for intercell interference nulling. This is despite the fact that (a) the BSs are densely deployed and the overall network is interference limited, and (b) the analysis does not account for the cost of channel estimation. Thus, even with perfect CSI available for all relevant channels at all BSs without any cost, to maximize the sum-rate, there is little gain in using interference nulling. When accounting for the cost of acquiring the required inter-cell CSI, it is unlikely that nulling interference at users in neighboring cells would bring considerable gain.

We emphasize that we do not claim that interference nulling is never useful. In particular, in trying to identify scenarios where it can provide a tangible benefit, we further show, via simulations, that nulling interference in neighboring cells can significantly improve cell-edge rates. This is particularly so when, for each user, the range of the cooperating cluster is chosen adaptively. In this paper, for each given user, this range is set in proportion to the distance to its serving BS. The implication of our work, therefore, is that inter-cell interference nulling should be reserved for those users who most need it.

\subsection{Related Work}

The trade-off between multiplexing and diversity for the single-user MIMO channel~\cite{ZT03} and for the single-cell multiuser uplink~\cite{TVZ04} are well known. However, the \emph{optimal} allocation of spatial resources between multiplexing and diversity in a multicell setting has not been studied rigorously. The multicell setting is of particular interest because of the possibility of mitigating intercell interference by coordinating transmissions across multiple BSs, usually known as interference coordination~\cite{LSCHMNS12}. For example, the problems of joint design of beamformers across multiple BSs to minimize transmit power~\cite{ZHG09,DY10} and to maximize SINR~\cite{VPW10,SBH10} have been studied.

The LS-MIMO system we consider has the number of users on the same order as the number of BS antennas. In this regard, it is very different from a massive MIMO system; in massive MIMO systems, each BS is equipped with an asymptotically large number of antennas and serves its scheduled users independently. In this asymptotic regime, the multiuser and intercell interference completely vanishes; intercell cooperation is, therefore, not required; indeed, interference is not an issue and matched filtering is adequate~\cite{M10,RPLLM13}. However, with a finite number of antennas, we cannot assume that the interference vanishes.  This paper deviates from the massive MIMO literature in that we explicitly account for the effect of interference cancellation (at all scheduled intra-cell users and some out-of-cell users) using zero-forcing (ZF) beamforming, and account for the residual interference at users from other transmissions.

In our user-centric clustering strategy, users request nulling from all BSs within some chosen range. The joint design of user-centric clusters and downlink beamformers for a fixed network topology has been studied in~\cite{DY14,HSBL13}. To account for the randomness in BS and user locations, we use tools from stochastic geometry~\cite{H12}, which has been subject of extensive studies in recent years for the analysis of single-antenna systems~\cite{ABG11}, multiple-antenna systems~\cite{DGBA12}, interference coordination in cellular systems~\cite{AH13}, heterogeneous small-cell networks~\cite{JXA10}, and a non-coherent joint transmission scheme~\cite{TSAJ14,NMH14} amongst others. Stochastic models for the analysis of user-centric clustering in wireless systems, where each BS serves a single user, have been introduced in~\cite{LJLH15,WCC15}. 

We carry out a stochastic analysis of an LS-MIMO system under user-centric clustering, including residual inter-cell interference, wherein each BS serves multiple users, and obtain the optimal allocation of spatial resources to multiplexing, diversity and interference nulling. In~\cite{ZH14}, the authors provide an analysis of the coverage probability including inter-cell interference coordination and intra-cell diversity for a single user; in contrast, we account for ZF for both spatial multiplexing and interference nulling.

\subsection{Summary of Contributions}

The main contributions of this paper are as follows:

\begin{enumerate}
\item\emph
{Stochastic analysis of LS-MIMO systems with user-centric clustering}: We present a stochastic analysis of an LS-MIMO system under user-centric clustering. Using tools from stochastic geometry, we obtain tractable expressions for both the cumulative distribution function (CDF) of user rates and the per-BS ergodic sum rate as functions of average cluster size. These expressions enable us to efficiently evaluate the performance of LS-MIMO systems with user-centric clustering under different system parameters.

\item\emph
{Optimizing allocation of spatial resources for multiplexing, diversity and interference nulling}: The analysis described above allows us to answer the central question of this paper on the optimal allocation of spatial dimensions. The analysis reveals that a close-to-optimum strategy, in sum-rate sense, is to use a fraction of spatial dimensions for multiplexing, the remaining dimensions for providing diversity, and leaving \emph{none} of the spatial dimensions for interference nulling.

\item\emph
{Interference nulling with fixed-range or range-adaptive strategies }: We illustrate numerically that interference nulling does have the potential to significantly enhance cell-edge user rates. We compare a fixed-range nulling strategy, in which the cooperative cluster size is the same for all users, to a range-adaptive nulling strategy, in which cell-edge users use larger clusters. Specifically, we show that interference nulling significantly improves the $10^{\mr{th}}$-percentile rate.

\end{enumerate}

The main implication of this work, therefore, is that BS cooperation, via interference nulling, should be used sparingly, in a user-specific manner.

\subsection{Paper Organization }

The rest of the paper is organized as follows. Section II presents the system model and states our assumptions. Section III provides the main analytic results, namely the ergodic rate characterizations of LS-MIMO systems with interference nulling. Section IV illustrates the benefit of interference nulling as a function of cluster size; the setting in this section requires that the number of BS antennas grows with cluster size. Section V answers the key question of this paper: given a fixed number of antennas per BS, how to optimize the number of users scheduled, the diversity order and dimensions for interference nulling to maximize sum-rate. Section VI focuses on cell-edge rates and explores the possibility of tailoring the cluster size to the user illustrating where cooperation provides a tangible benefit. Finally, we wrap up the paper with some conclusions in Section VII.

\subsection{Notation}

Lower-case letters, e.g., $g$, denote scalars, bold-face lower-case letters, e.g., $\g$, denote vectors, and bold-face upper-case letters, e.g., $\G$, denote matrices. We use $\I_n$ to denote an $n \times n$ identity matrix, $\LB \cdot\RB\htp$ to denote the Hermitian and $\G^{\dagger}=(\G\htp \G)^{-1}\G\htp$ to denote the left pseudo-inverse of $\G$. We use $\CC$ to denote the set of complex numbers. The complex Gaussian distribution with mean $\m$ and covariance matrix $\bm{\Sigma}$ is denoted by $\mathcal{CN}(\m,\bm{\Sigma})$, and $\Gamma \LB k,\theta\RB$ denotes a Gamma distribution with shape parameter $k$ and scale parameter $\theta$; finally, $\EE [\cdot]$ denotes statistical expectation, $\| \cdot\|_2$ denotes the $\ell_2$ norm of a vector and $\lv \cdot \rv$ denotes the cardinality of a set.

\section{System Model} \label{sec:sys_model_chap4}

Consider a MIMO system where BSs are distributed according to a homogeneous PPP $\Phi$ with density $\lambda$ over the entire $\RR^2$ plane. Each BS is equipped with $M$ antennas, and is constrained to use a maximum power of $P_T$. Single-antenna users are distributed according to an independent homogeneous PPP with density considerably larger than that of the BSs. Users form cells by associating with their closest BSs, thereby partitioning $\RR^2$ into a Voronoi tessellation. Perfect CSI is assumed to be available at the BSs and the users.

The wireless fading channel is modeled as follows: let $\g_{ij} = \sqrt{\beta_{ij}} \h_{ij} \in \Cc^{M\times 1}$ denote the channel between BS $j$ and user $i$, where $\h_{ij} \sim \Cc \Nc \LB \zerov,\I_{M} \RB$ indicates the small-scale Rayleigh fading component and $\beta_{ij} = \LB 1 + \frac{r_{ij}}{d_o} \RB^{-\alpha}$ is the path-loss component; here, $r_{ij}$ is the distance between user $i$ and BS $j$, $\alpha > 2$ denotes the path-loss exponent and $d_o$ the reference distance.

In a given time-frequency slot, we assume that the BS $b$ schedules a set of $K$ users, denoted by $\Sc_b$, chosen randomly from its associated users according to a round-robin scheduling scheme.\footnote{This model implies that while the point processes associated with the BSs and \emph{potential} users are independent, the density of the \emph{scheduled} users is not uniform. Smaller cells have a higher density of scheduled users and vice versa.} The BS uses ZF beamforming with equal power allocated across the $K$ users. Further, BS $b$ uses its available antennas to cancel interference at a subset of \emph{out-of-cell} users, denoted by $\Oc_b$. The normalized ZF beam, assigned to user $i$ associated with BS $b$, while nulling the interference at the other $K -1 + |\Oc_b|$ users, is given by
\begin{equation}
\w_{ib} = \frac{\LB \I_{M} - \G_{-ib}\G_{-ib}^\dagger \RB \hat{\g}_{ib}}{\Big \| \LB \I_{M} - \G_{-ib}\G_{-ib}^\dagger \RB \hat{\g}_{ib} \Big\|_2},~\forall{i} \in \Sc_{b}
\end{equation}
where  $\G_{{-i}b} = \LSB \hat{\g}_{1b},\ldots, \hat{\g}_{\LB i-1\RB b},\hat{\g}_{\LB i+ 1\RB b}, \ldots,\hat{\g}_{\LB K + \lv \Oc_b\rv \RB b}\RSB$ and $\hat{\g}_{ib} = \g_{ib}/ \| \g_{ib} \|_2$.\footnote{Note that in an LS-MIMO system, since each BS independently serves its own scheduled users, data sharing across BSs over backhaul is not required. Further, to perform inter-cell interference nulling, each BS requires access to the CSI between itself and out-of-cell users. Hence, CSI exchange across BSs is also not required. }

We note that although ZF beamforming completely eliminates interference at specific users, it does require substantial CSI and is suboptimal as compared to, e.g., minimum-mean squared error (MMSE) beamforming. In the absence of complete CSI, it may even be inferior to maximum
ratio transmission (MRT), especially in the massive MIMO regime, where MRT is
near optimal when the number of BS antennas is much larger than the number of
users. Nevertheless, this paper chooses to investigate ZF, because we are
interested in the regime where the number of users being served is a
substantial fraction of the number of BS antennas and also because ZF offers a
reasonable balance between performance and analytic tractability.

\subsection{User-Centric Clustering in LS-MIMO Systems} \label{sec:simul}

The choice of out-of-cell users for interference cancellation depends on the BS clustering strategy. This paper considers \emph{user-centric clustering}, in which each user forms a cooperative cluster comprising its serving BS and a set of neighboring BSs from which to request interference nulling. Specifically, each user requests interference nulling from all BSs within a range of $R_c$, which, given our channel model, is equivalent to each user requesting nulling from BSs with average interference power above a chosen threshold.  Ideally, interference from all the BSs within the cluster would be cancelled; in this case, only the interference from outside of the cluster remains and is treated as additional noise. However, in reality, because each BS has finite number of available spatial dimensions, it can only select a subset of out-of-cell users for interference nulling. In particular, each BS reserves sufficient antennas to serve its $K$ scheduled users with at least diversity order $\zeta$. Accounting for the orthogonal property of ZF beamforming, BS $b$ must reserve $\zeta + K - 1$ dimensions for transmission to its own users, leaving a maximum of
\begin{equation}
O = \lv \Oc_b\rv = M - K - \zeta + 1,~\forall b \label{eq:max_user}
\end{equation}
dimensions for interference nulling at out-of-cell users. If the number of users requesting interference nulling at BS $b$ is larger than $O$, then BS $b$ selects the $O$ users with the strongest channel magnitudes within the candidate set, i.e., some out-of-cell users' requests are not granted. However, if the number of users requesting interference cancellation from BS $b$ is smaller than $O$, then BS $b$ cancels interference at all out-of-cell users and uses the extra antennas to provide intra-cell users with a diversity order larger than $\zeta$.

In the system described above, let $\Phi_{S_i} \subset \Phi$ be the set of BSs located in the cluster chosen by user $i$, and $\Phi_{I_i} = \Phi \setminus \Phi_{S_i}$ be the set of interfering BSs located outside the cluster. Further, let $\Phi_{S_i,\mr{Intf}} \subset \Phi_{S_i}$ indicate the BSs in $\Phi_{S_i}$ that do not have sufficient antennas to cancel interference at user $i$. The received signal at user $i$ associated with BS $b$ is the sum of the intended signal, residual intra-cluster interference, inter-cluster interference, and receiver noise; the received signal is, therefore, given by
\begin{multline}
y_{ib} = \underbrace{\sqrt{\frac{P_T}{K}} \g_{ib}\htp \w_{ib}
s_{ib}}_{\text{intended signal}} +  \underbrace{\sum_{j \in
\Phi_{S_{i},\mr{Intf}}} \sum_{k = 1}^{K} \sqrt{\frac{P_T}{K}} \g_{ij}\htp
\w_{kj} s_{kj}}_{\text{intra-cluster interference}} \\
+  \underbrace{\sum_{m \in
\Phi_{I_i}} \sum_{u = 1}^{K} \sqrt{\frac{P_T}{K}} \g_{im}\htp \w_{um}
s_{um}}_{\text{inter-cluster interference}} +
\underbrace{n_{ib}}_{\text{noise}} \label{eq:3}
\end{multline}
where $s_{ib}$ denotes the complex symbol intended for user $i$ associated with BS $b$ such that $\EE \LSB \lv s_{ib}\rv^2 \RSB = 1$, and $n_{ib}$ denotes the additive white Gaussian noise with variance $\sigma^2$.

In this system model, each BS chooses $K$ users out of all the potential users in the cell. Since the user and BS distributions are independent, different cells are likely to have a different number of potential users, denoted by $K_b$ in cell $b$. Thus, in practice, each set of $K$ scheduled users would be served $K/K_b$ fraction of the time which varies across cells. In the paper, we assume that this fraction is fixed and identical across the network. We emphasize that this assumption does not affect the achievable sum rate. In Section~\ref{subsec:normalized}, we confirm numerically that this assumption does not materially change the characterization of the downlink rate distribution either.

\subsection{Analytical Model of User-Centric Clustering}
\label{sec:analytic_model}

This paper aims to provide a performance analysis of the user-centric LS-MIMO system described above. To enable a tractable theoretical analysis, we adopt the following simplifying assumptions. The validity of these assumptions is evaluated via simulations later in this paper.
\begin{enumerate}


\item[$A1:$] We assume that the number of BSs within a range of $R_c$ from each user is equal to the average number of BSs within the area according to the density of the PPP, i.e., the circle of radius $R_c$ centered at a user has exactly $\bar{B} = \left| \Phi_{S_i} \right| = \lambda \pi R_c^2$ BSs $\forall i$.

\item[$A2:$] Since users request interference nulling from BSs within a range of $R_c$, each BS must cancel interference at all users located within distance $R_c$ from itself. We assume that each BS has enough antennas to serve its $K$ scheduled users with a fixed diversity order of $\zeta$, and to grant all interference nulling requests. Essentially, \emph{the analysis} assumes \emph{complete} intra-cluster interference cancellation using as many spatial dimensions as the \emph{average} number of interference nulling requests. We, therefore, drop the second term in (\ref{eq:3}). 
\end{enumerate}

In the first part of this paper, we assume that the cluster radius $R_c$ is the
same for all the users. In this case, from $A2$, each BS will cancel interference
at $O = M - K - \zeta + 1$ out-of-cell users, while serving its $K$ scheduled
users with diversity order of $\zeta$. To derive a relationship between $R_c$, $\bar{B}$, $O$ and $K$, consider an area 
comprising $N$ BSs. The total number of
scheduled users is $NK$. The total number of dimensions available at the $N$
BSs to serve these users and to null interference is $(O+K)N$. Therefore, the
average number of BSs within each user's cluster must be: 
\begin{equation}
\bar{B} = \LB O + K \RB/K.
\label{eq:average_B}
\end{equation}
Now based on $A1$, since there are exactly $\bar{B}$ BSs within the cluster radius, we must have
\begin{equation}
R_c = \sqrt{\bar{B}/\lambda \pi}.
\end{equation}



Under these assumptions, using (\ref{eq:3}), the signal-to-interference-plus-noise ratio (SINR) of user $i$ scheduled by BS $b$ can now be expressed as
\begin{equation}
\gamma_{ib} = \frac{\rho \beta_{ib}\lv \h_{ib}\htp \w_{ib}\rv^2}{\sum_{m \in \Phi_{I_i}} \sum_{k = 1}^{K} \rho \beta_{im} \lv \h_{im}\htp \w_{km}\rv^2 + 1} \label{eq:sinr_chap4}
\end{equation}
where $\rho = \frac{P_T}{K\sigma^2}$ indicates the per-user signal-to-noise ratio (SNR). The effective instantaneous achievable rate of user $i$, in bit/s/Hz, denoted by $R_{ib}$, is therefore given by
\begin{equation}
R_{ib} = \log_2 \LB 1+ \frac{\gamma_{ib}}{\tau} \RB
\end{equation}
where $\tau$ represents the SNR gap to capacity.

\section{Stochastic Geometry Analysis of LS-MIMO}
\label{sec:performance_chap4}

The main analytic results of this paper are a characterization of an upper bound on the complementary cumulative distribution function (CCDF) of user rates and a characterization of per-BS ergodic sum rate for LS-MIMO systems with user-centric clustering using stochastic geometry methods. These novel analyses substantially advance previous stochastic analysis of inter-cell interference coordination systems \cite{ZH14} in analyzing multiuser transmissions and explicitly taking into account the effect of ZF beamforming for both spatial multiplexing and interference nulling.

\begin{theorem}\label{ub_ccdf}
Under assumptions $A1$-$A2$, for an LS-MIMO system with user-centric clustering and a constant cluster size for each user, providing a fixed diversity order $\zeta$ for each scheduled user, and employing ZF beamforming at the BSs, with equal power across the downlink beams, the CCDF of the downlink user rates satisfies
\begin{multline}
\PP \LB w R_{ib} \geq \kappa \RB \leq  \int_{r = 0}^{\infty} \sum_{j =
1}^{\zeta} \binom{\zeta}{j } \LB -1 \RB^{j + 1} e^{-s} \\
\exp \LB 2\pi\lambda \LB \Psi_{\mr{I}}(s) - \Psi_{\mr{II}}(s) \RB \RB
\mr{d}F_{r_{\mr{min}}}(r), \label{eq:ub_ccdf}
\end{multline}
where
\begin{equation}
F_{r_{\mr{min}}}(r) = 1 - \exp \LB -\lambda \pi r^2 \RB,~r>0
\label{F_r}
\end{equation}
is the CDF of distance between a user and its closest BS, $s$ is given as
\begin{equation}
s = \LB \zeta!\RB^{\frac{-1}{\zeta}} j \frac{\tau}{\rho}\LB 2^{\frac{\kappa}{w}} - 1 \RB \LB 1 +
\frac{r}{d_0} \RB^{\alpha}
\end{equation}
where $w$ accounts for the fraction of the available time allocated to each set of $K$ users and $\Psi_{\mr{I}}(s)$ and $\Psi_{\mr{II}}(s)$ are given by
\begin{multline}
\Psi_{\mr{I}}(s) = \frac{d_0^2}{2}\LB 1 + \frac{R_c}{d_0}\RB^2 \\
\LB 1 - {}_2 F_1 \LB K,-\frac{2}{\alpha};1 - \frac{2}{\alpha}; -s\rho \LB 1 + \frac{R_c}{d_0}\RB^{-\alpha} \RB \RB \label{eq:psi1}
\end{multline}
and
\begin{multline}
\Psi_{\mr{II}}(s) = d_0^2 \LB 1 + \frac{R_c}{d_0}\RB \\
\LB 1 - {}_2 F_1 \LB K,-\frac{1}{\alpha};1 - \frac{1}{\alpha}; -s\rho \LB 1 + \frac{R_c}{d_0}\RB^{-\alpha} \RB \RB. \label{eq:psi2}
\end{multline}
\end{theorem}

\begin{IEEEproof}
In order to prove the theorem, we use the Alzer's inequality~\cite{LJLH15} which states that the CCDF of a gamma random variable $X \sim \Gamma \LB m,1\RB$ satisfies
\begin{equation}
\PP \LB X \geq \kappa \RB \leq 1 - \LB 1 - e^{-\mu \kappa} \RB^m \nonumber
\end{equation}
with $\mu = \LB m!\RB^{\frac{-1}{m}}$.

As the first step, let 
\[I_{ib} = \sum_{m \in \Phi_I} \sum_{k = 1}^{K} \rho \LB 1 + \frac{r_{im}}{d_0} \RB^{-\alpha} \lv \h_{im} \htp \w_{km} \rv^2 \] be the aggregate inter-cluster interference power seen at user $i$ associated with BS $b$. Considering a round-robin scheduling scheme, the rate CCDF of user $i$ is given by
\begin{subequations}
\label{eq:inequality}
\begin{align}
\PP& \LB \gamma_{ib} \geq \tau \LB 2^{\frac{\kappa}{w}} - 1 \RB \RB \nonumber \\
& =\PP \LB \lv \h_{ib} \htp \w_{ib} \rv^2 \geq \frac{\tau\LB 2^{\frac{{\kappa}}{w}} - 1 \RB \LB 1 + \frac{r_{ib}}{d_0}\RB^{\alpha} }{\rho} \LB I_{ib} + 1 \RB\RB  \\
& \leq \int_{0}^{\infty} \frac{\mr{d} F_{r_{\mr{min}}}}{\mr{d}r} 
 \EE_{I_{ib}} \LSB 1 - \LB 1 - 
\vphantom{\exp \LB -\frac{\mu\tau \LB 2^{\frac{{\kappa}}{w}} \RB^2}{\rho} \RB}
\right. \right. \nonumber \\
& \quad \left. \left.  \exp \LB -\frac{\mu\tau \LB 2^{\frac{{\kappa}}{w}} - 1 \RB \LB 1 + \frac{r}{d_0} \RB^{\alpha}}{\rho} \LB I_{ib} + 1 \RB \RB \RB^{\zeta} \RSB \label{eq:7c}. 
\end{align}
\end{subequations}
Given that $\lv \h_{ib}\htp \w_{ib} \rv^2 \sim \Gamma \LB \zeta,1 \RB$~\cite{HYA15}, the inequality is obtained by first conditioning on the interference power, and then using Alzer's inequality where $\mu = \LB \zeta! \RB^{\frac{-1}{\zeta}}$.

Now, let $s = \mu j \tau \LB 2^{\frac{\kappa}{w}} - 1\RB \LB 1 + \frac{r}{d_0} \RB^{\alpha}/\rho$. Using the binomial expansion, the expectation in (\ref{eq:7c}) can be written as
\begin{equation}
\sum_{j = 1}^{\zeta} \binom{\zeta}{j } \LB -1 \RB^{j + 1} e^{-s} \Lc_{I_{ib}}\LB s \RB \label{eq:binomial}
\end{equation}
where 
$\Lc_{I_{ib}}\LB s \RB = \EE_{I_{ib}}  \LB e^{-s I_{ib}} \RB$ denotes the Laplace transform of the probability density function of the inter-cluster interference power at user $i$. The final part of the proof therefore is to obtain $\Lc_{I_{ib}}\LB s \RB $, which can be carried out as follows:
\begin{align}
&\Lc_{I_i} \LB s \RB = \EE_{\Phi_I,\h} \LSB e^{-s \sum_{m \in \Phi_I} \sum_{k=1}^{K} \LB 1 +  \frac{r_{im}}{d_0} \RB^{-\alpha} \lv \h_{im} \htp \w_{km}\rv^2}\RSB \nonumber \\
&\stackrel{\LB a\RB}{=} \EE_{\Phi_{I_i}} \LSB \prod_{m\in \Phi_I} \EE_{\h} \LSB e^{-s \LB 1 + \frac{r_{im}}{d_0}\RB^{-\alpha} \sum_{k=1}^{K}\lv \h_{im}\htp \w_{km}\rv^2} \RSB \RSB \nonumber \\
&\stackrel{\LB b\RB}{=}  \EE_{\Phi_{I_i}} \LSB \prod_{m \in \Phi_I} \LB 1 + s\LB 1 + \frac{r_{im}}{d_0}\RB^{-\alpha} \RB^{-K}\RSB \nonumber \\
&\stackrel{\LB c\RB}{=} \exp \LB -2\pi\lambda  \int_{R_c}^{\infty} \LSB 1 - \LB 1 + s\rho \LB 1 + \frac{v}{d_0}\RB^{-\alpha}\RB^{-K} \RSB v \mr{d}v \RB\nonumber \\
&\stackrel{\LB d\RB}{=} \exp \LB 2\pi\lambda \LB \Psi_{\mr{I}} \LB s \RB - \Psi_{\mr{II}} \LB s \RB \RB \RB \label{eq:exp_func}
\end{align}
where $\LB a \RB$ follows from the fact that the small-scale channel fading is independent of the BS locations. Similar to the approach taken in other works, e.g.~\cite{DGBA12,DKA13}, for the analysis, we assume that users scheduled by each interfering BS have orthogonal channels. The aggregate interference power imposed by each interfering BS at user $i$ is therefore distributed as $\sum_{k = 1}^{K} \lv \h_{im}\htp \w_{km}\rv^2 \sim \Gamma \LB K,1\RB. $
Relation $\LB b\RB$ is then obtained using the moment generating function (MGF) of this Gamma distribution. To derive $\LB c \RB$, we use the probability generating functional of a PPP with density $\lambda$ given as~\cite{H12} 
\begin{equation}
\EE \LB \prod_{\x \in \Phi} v \LB \x \RB \RB = \exp \LB -\lambda \int_{\RR^2} \LB 1 - v\LB \x \RB \RB \mr{d}\x\RB \label{eq:mgf_ppp}
\end{equation}
and change the coordinates from Cartesian to polar. Note that the region of integration is $[R_c,\infty)$ since interfering BSs are located at distance at least $R_c$ away from user $i$. Finally, $\LB d\RB$ follows by using standard integral formulae for the $_2F_1(\cdot, \cdot; \cdot; \cdot)$ function.

Inserting~\eqref{eq:exp_func} and~\eqref{eq:binomial} in~\eqref{eq:inequality} completes the proof.

\end{IEEEproof}

While Theorem~\ref{ub_ccdf} characterizes the CCDF of the user rates, our next theorem characterizes the per-BS ergodic sum rate.

\begin{theorem} \label{ergodic_rate_chap4}
Under assumptions $A1$ and $A2$, for an LS-MIMO system with user-centric clustering and constant cluster size for each user, providing a fixed diversity order $\zeta$ for each scheduled user, and employing ZF beamforming at the BSs with equal power across the downlink beams, the per-BS ergodic sum rate in nats/sec/Hz is given by
\begin{multline}
R_{cell} = K \int_{0}^{\infty} \frac{e^{-z \tau}}{z}
\exp \LB 2\pi \lambda \LB \Psi_{\mr{I}}\LB z \tau \RB - \Psi_{\mr{II}} \LB z\tau\RB \RB \RB \\
\LB 1 - \int_{0}^{\infty} \frac{\mr{d} F_{r_{\mr{min}}}(r)}{\mr{d}r} \LB 1 + z\rho \LB 1 + \frac{r}{d_0}\RB^{-\alpha} \RB^{-\zeta}\mr{d}r\RB \mr{d}z\label{eq:per_cell_sum_rate}
\end{multline}
where $F_{r_{\mr{min}}}(\cdot)$, $\Psi_{\mr{I}} \LB \cdot \RB$ and
$\Psi_{\mr{II}} \LB \cdot \RB$ are given in (\ref{F_r}) and
(\ref{eq:psi1})-(\ref{eq:psi2}).
\end{theorem}

\begin{IEEEproof}
The per-BS ergodic sum rate of an LS-MIMO system where each BS schedules $K$ users simultaneously is given by
\begin{eqnarray}
R_{cell} & = & K \EE_{\Phi,\h} \LSB \log_2 \LB 1 + \frac{\gamma_{ib}}{\tau} \RB \RSB \nonumber \\
& = & K \EE_{\Phi,\h} \LSB \int_{0}^{\infty} \frac{e^{-t}}{t} \LB 1 - e^{-\gamma_{ib}t} \RB \mr{d}t \RSB \label{eq:rcell}
\end{eqnarray}
where the second expression uses an alternate form of the rate function~\cite{H08} given as
\begin{equation}
\log \LB 1 + x \RB = \int_{0}^{\infty} \frac{e^{-t}}{t} \LB 1 - e^{-xt} \RB \mr{d}t. \nonumber
\end{equation}
Let  \[I_{ib} = \sum_{m \in \Phi_{I_i}} \sum_{k = 1}^{K} \rho \LB 1 + \frac{r_{im}}{d_0} \RB^{-\alpha} \lv \h_{im} \htp \w_{km} \rv^2\] and $t = z \tau \LB I_{ib} + 1 \RB$. The expression in~\eqref{eq:rcell} can be simplified as
\begin{subequations}
\label{eq:cell_rate}
\begin{align}
R_{cell} = \ & \EE_{\Phi,\h} \LSB \int_{0}^{\infty} \frac{e^{-z \tau}}{z}  \exp \LB -z \tau I_{ib} \RB \right. \nonumber \\
 \ & \left. \vphantom{\int} \qquad \LB 1 - \exp \LB -z \rho \beta_{ib} \lv \h_{ib} \htp \w_{ib} \rv^2 \RB \RB \mr{d} z \RSB \\
= \ & \int_{0}^{\infty} \frac{e^{-z \tau}}{z} \EE_{\Phi_{I_i},\h} \LSB  \exp \LB -z \tau I_{ib} \RB \RSB \nonumber \\
 \ & \quad \LSB 1 - \EE_{\Phi_{S_i},\h} \LB \exp \LB -z \rho \beta_{ib} \lv \h_{ib} \htp \w_{ib} \rv^2 \RB \RB \RSB \mr{d} z \label{eq:cell_rate1} 
\end{align}
\end{subequations}
where based on the Fubini theorem, the order of expectation and integration operations are exchanged. Further, $\Phi_{I_i}$ and $\Phi_{S_i}$ are statistically independent. Therefore,~\eqref{eq:cell_rate1} follows.

In~\eqref{eq:cell_rate1}, $\EE_{\Phi_{I_i},\h} \LSB  \exp \LB -z \tau I_{ib} \RB \RSB = \Lc_{I_{ib}} \LB s\RB \lv_{s = z \tau}$ is the Laplace transform of the density function of the aggregate inter-cluster interference power at user $i$, and is given in~\eqref{eq:exp_func}. Further, the expectation of the signal term can be computed as
\begin{align}
& \EE_{\Phi_{S_i},\h} \LB \exp \LB -z \rho \LB 1 + \frac{r_{ib}}{d_0} \RB^{-\alpha} \lv \h_{ib} \htp \w_{ib} \rv^2 \RB \RB \nonumber \\
&\stackrel{\LB a\RB}{=}   \EE_{\Phi_{S_i}} \LSB \EE_{\h} \LB\exp \LB -z \rho \LB 1 + \frac{r_{ib}}{d_0} \RB^{-\alpha} \lv \h_{ib} \htp \w_{ib} \rv^2 \RB \RB \RSB \nonumber \\
&\stackrel{\LB b\RB}{=} \EE_{\Phi_{S_i}} \LSB \LB 1 + z \rho \LB 1 + \frac{r_{ib}}{d_0} \RB^{-\alpha} \RB^{-\zeta} \RSB \nonumber \\
&\stackrel{\LB c\RB}{=} \int_{0}^{\infty} \frac{\mr{d}F_{r_{\mr{min}}}\LB r\RB}{\mr{d}r} \LB 1 + z\rho \LB 1 + \frac{r}{d_0} \RB^{-\alpha} \RB^{-\zeta}. 
\end{align}
In $\LB a\RB$, we used the fact that $\Phi_{S_i}$ and $\h$ are statistically independent. The signal power is distributed as $\lv \h_{ib}\htp \w_{ib}\rv^2 \sim \Gamma \LB \zeta,1\RB$. Hence, $\LB b\RB$ is obtained from the MGF of a Gamma distribution. Finally, $\LB c\RB$ is derived by noting that $r_{ib}$ is the distance between user $i$ and its closest BS.

Inserting the interference and signal terms in~\eqref{eq:cell_rate1} completes the proof.

\end{IEEEproof}


We emphasize that the expression in Theorem~\ref{ergodic_rate_chap4} is not derived directly from Theorem~\ref{ub_ccdf}, but by using the more tractable form of the rate expression~\cite{H08}
\begin{equation}
\log \LB 1 + x \RB = \int_{0}^{\infty} \frac{e^{-t}}{t} \LB 1 - e^{-xt} \RB \mr{d}t \nonumber
\end{equation}
and substituting in the Laplace transforms of the density functions of the interference and signal powers. The proof uses the same approach as in~\cite{LY14,HYA15,LJLH15}.

\begin{rem}
Theorem~\ref{ub_ccdf} and Theorem~\ref{ergodic_rate_chap4} completely characterize the CCDF of the user rates and the per-BS ergodic sum rate of an LS-MIMO system with user-centric clustering as a function of important system parameters, such as BS deployment density ($\lambda$), number of users scheduled by each BS ($K$), diversity order per user ($\zeta$), and cluster radius ($R_c$). 
\end{rem}
\begin{rem}
Although the expressions appear complicated, they are nevertheless can be computed more efficiently as compared to performing system level simulations. In particular, although the analytical expressions involve double integrals, using the transformation introduced in Remark $1$ of~\cite{ZH14}, they can be computed efficiently.
\end{rem}

\section{Benefits of Interference Nulling as Function of Cluster Size}

The LS-MIMO system considered in this paper uses spatial dimensions at the BSs to mitigate intercell interference. It is therefore expected that system performance improves as more BSs are included in the cooperating cluster for each user. This section uses the analytic expressions derived in the previous section to quantify the performance improvement of LS-MIMO systems as a function of cluster size. To do this, we fix the number of users scheduled in each cell and the diversity order for each user, and let the cluster size increase. In this case, to allow for interference nulling, a larger cluster size also implies a larger number of antennas at each BS. Thus, the system performance benefits characterized in this section are largely due to the increase in the number of BS antennas. These results also serve to validate the accuracy of the analytic expressions developed in Section III. 

We further compare the performance of LS-MIMO systems with that of conventional single-cell processing. Note that this section does not yet address the question of the optimal choice of the number of users to schedule, the diversity order to provide, and the amount of interference nulling to do, with a fixed number of BS antennas. The answer to this important question is deferred to the next section. 

The system parameters for the numerical simulation of this section are listed in Table~\ref{table:parameters}. Each BS schedules $K = 3$ users, and provides diversity order of $\zeta = 4$ for each of its scheduled users. In a conventional system without interference nulling, this can be achieved with each BS equipped with $M = 6$ transmit antennas. In the MIMO system, for a fixed average cluster size $\bar{B}$, each BS is equipped with $M = \zeta + K\bar{B} -1$ transmit antennas in order to null interference at $O = K \LB \bar{B} - 1 \RB$ out-of-cell users. All BSs share the available bandwidth of $W = 20$ MHz with frequency reuse factor of one. Here, $K_b = 21,~\forall b$, i.e., each set of $K$ users is scheduled $w= K/K_b = 1/7$ of the time. The analytical results are obtained using Theorem~\ref{ub_ccdf}.

\begin{table}[]
\caption{System Design Parameters}
\centering
    \begin{tabular}{|c| c|}
    \hline
    BS density & $\lambda = 1/\pi 500^2$ m$^{-2}$\\ \hline
    Total bandwidth & $W = 20$ MHz \\ \hline
    BS max. available power & $43$ dBm \\ \hline
    Background noise PSD & $N_o = -174$ dBm/Hz \\ \hline
    Noise figure & $N_f = 9$ dB \\ \hline
    SNR gap to capacity& $\tau = 3$ dB \\ \hline
    Path-loss exponent & $\alpha =3.76$ \\ \hline
    Reference distance & $d_0 = 0.3920$ m \\ \hline
         \end{tabular}
    \label{table:parameters}
\end{table}

\begin{figure}[t!]
\centering
\includegraphics[scale=1.05]{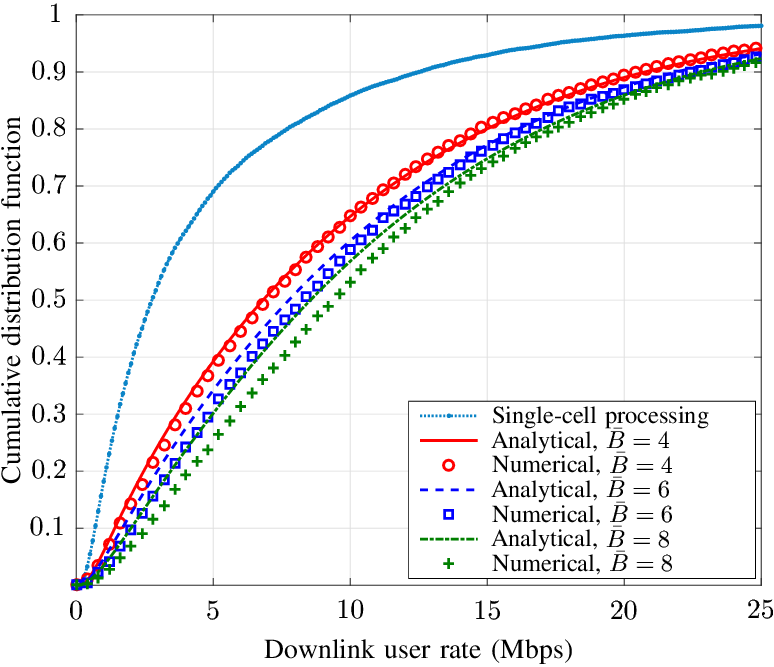}
\caption{CDF of the downlink user rates in a conventional system (with $M =6$, $K=3$ and $\zeta = 4$), and LS-MIMO systems with user-centric clustering (with $\bar{B} = 4$, $6$, and $8$, $K=3$, $\zeta = 4$, and $M = \zeta + K\bar{B} - 1$).}
\label{fig:increasing_antenna}
 \end{figure}

Fig.~\ref{fig:increasing_antenna} plots the downlink rate distributions of various LS-MIMO systems with different cluster sizes, averaged over the network topology and realizations of channel fading. The figure compares the analytic results using the expression in (\ref{eq:ub_ccdf}) with numerical results obtained via simulations for an LS-MIMO system as described in Section~\ref{sec:simul} (i.e., without assuming A1 or A2). The CDF of the downlink user rate associated with single-cell processing is obtained via simulation. 

As is clear from the figure, the theory developed in Section III is accurate despite the simplifying assumptions made. The biggest source of the small discrepancy between the analytical and numerical results is the assumptions A1 and A2, made in developing the analysis, that intra-cluster interference within the cluster radius is \emph{completely} eliminated, while in reality, this may not be always true.

As shown in Fig.~\ref{fig:increasing_antenna}, increasing the average cluster size significantly improves the downlink user rates. For example, the MIMO system with $\bar{B} = 8$ provides a $50\%$ rate improvement for the $20^{\mr{th}}$-percentile users as compared to the MIMO system with $\bar{B} = 4$, which already doubles the $20^{\mr{th}}$-percentile user rate as compared to the single-cell processing baseline. We emphasize that these benefits are possible because of the significant increase in the number of BS antennas as cluster size increases.


\section{Diversity, Multiplexing or Interference Cancellation?}
\label{sec:numerical2_chap4}

It is clear from results presented in the previous section that enabling interference nulling by deploying more antennas at each BS significantly improves the achievable rate of all users.~\emph{However, is interference nulling the best use of the available spatial degrees-of-freedom?} After all, the excess number of BS antennas could also have been used to schedule more users or to provide a larger diversity gain for the scheduled users. What is the optimal trade-off among the three?

In this section, we aim to answer the central question of this paper: given a \emph{fixed} number of antennas ($M$) at each BS, what are the optimal number of users to schedule ($K$). and the optimal number of interference directions to null ($O$), to maximize the per-BS ergodic sum rate? (Note that only two of the triple $(K,\zeta,O)$ are free variables.)

The per-BS ergodic sum rate expression (obtained in Theorem~\ref{ergodic_rate_chap4}) in Section~\ref{sec:performance_chap4} gives us an efficient way to answer this question. Keeping $M$ constant, we can obtain the operating point that maximizes the sum-rate of an LS-MIMO system by choosing the optimal triple $(K,\zeta, O)$ with $K \in \{ 1,\cdots,M\}$, $O \in \{ 0,\cdots,M - K \}$ and $\zeta = M - K - O + 1$. In the following, we present both theoretical results and numerical results obtained via simulations for various choices of $(K,\zeta,O)$.  Numerical simulations are averaged over network topologies and small-scale channel fading realizations. The system parameters are again as listed in Table~\ref{table:parameters}. Here the total number of users associated with each BS is $K_b = 15,~\forall{b}$. 

\begin{figure}[t!]
\centering
\includegraphics[scale=1.05]{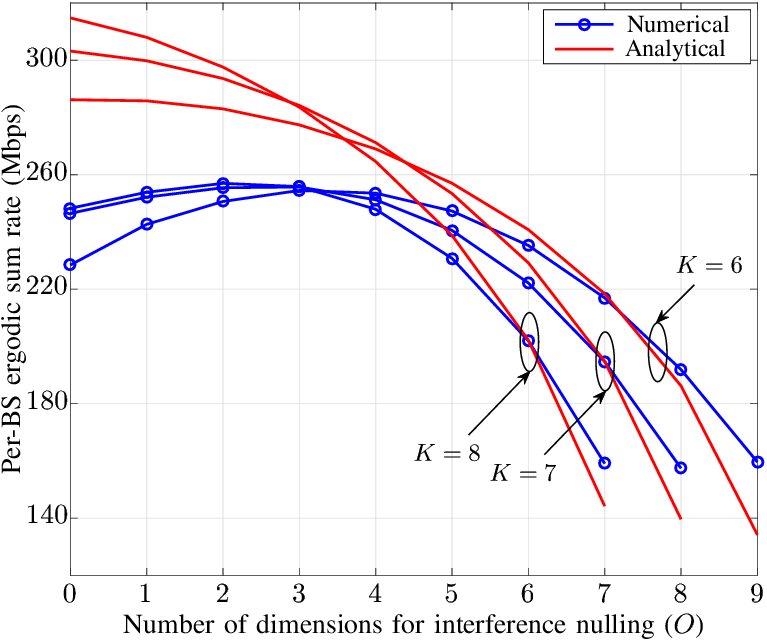}
\caption{Per-BS ergodic sum rate of an LS-MIMO system with $M = 15$, $K = 6$, $7$, $8$, and various choices of $\zeta$ and $O$ obtained both numerically and analytically. }
\label{fig:M15_optimum1}
 \end{figure}

\begin{figure}[t!]
\centering
\includegraphics[scale=1.05]{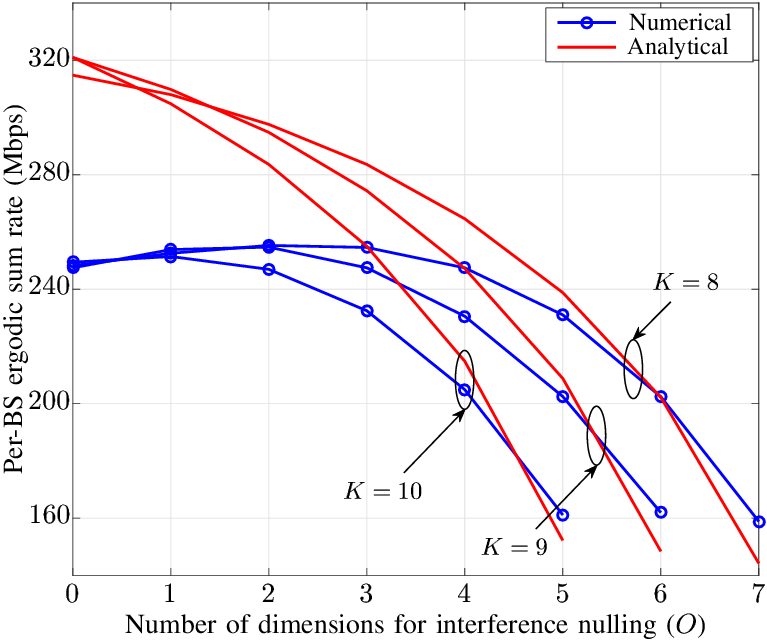}
\caption{Per-BS ergodic sum rate of an LS-MIMO system with $M = 15$, $K = 8$, $9$, $10$, and various choices of $\zeta$ and $O$ obtained both numerically and analytically. }
\label{fig:M15_optimum2}
 \end{figure}

Figs.~\ref{fig:M15_optimum1} and~\ref{fig:M15_optimum2} plot the per-BS ergodic sum rate of an LS-MIMO system with $M = 15$ for various combinations of $\LB K,\zeta,O\RB$ obtained analytically using (\ref{eq:per_cell_sum_rate}) and numerically via simulation for the system model described in Section~\ref{sec:simul} (i.e., without assumptions A1 or A2). First, observe that when a substantial number of dimensions are devoted to nulling interference (large $O$), the numerical and analytic results are in good agreement. However, at low values of $O$, the two results diverge somewhat. The main reason for the discrepancy is that, in the simulations, once $O$ is specified, the \emph{exact} number of BSs nulling interference at each user, hence its exact cluster size, can be  determined given the network topology. The cluster size may also vary from one user to the other. However, as explained in Section~\ref{sec:analytic_model}, in the analysis, the values of $K$ and $O$ specify the \emph{average} number of BSs in a cluster, i.e., $\bar{B} = \LB K + O \RB /K$. Despite the discrepancy, the analysis does capture the general behavior of the system. For example, when both $O$ and $K$ are small, increasing $K$ improves the per-BS ergodic sum rate, while for large values of $O$ and $K$, increasing $K$ degrades system performance. However, more importantly, the analysis helps identify the region of $O$ which provides the maximum per-BS ergodic sum rate, leading to the main, and surprising, result of this paper.


Our analytical and simulation results reveal a remarkable trend: the per-BS ergodic sum rate is largely a \emph{decreasing} function of the number of spatial dimensions assigned to null interference. The analytic expression in (\ref{eq:per_cell_sum_rate}) is, in fact, a strictly decreasing function of $O$, while the numerical curves are approximately so. It is, therefore, nearly optimum for each BS to operate independently of the other BSs and \emph{not to spend any spatial resources on nulling interference}. Specifically, the optimal operating point, in sum-rate sense, obtained numerically is $\LB K^*,\zeta^*,O^*\RB = \LB 8,6,2\RB$, and the optimal operating point obtained analytically is $\LB K^*,\zeta^*,O^*\RB = \LB 10,6,0\RB$. The difference between the achievable sum rates at these two operating points, when obtained numerically via simulations, is only about $4\%$. Thus, when $K$ and $\zeta$ are properly chosen, an LS-MIMO system \emph{without interference nulling}, i.e., with each BS operating independently, is close to being \emph{sum-rate} optimal.

\begin{figure}[t!]
\centering
\includegraphics[scale=0.36]{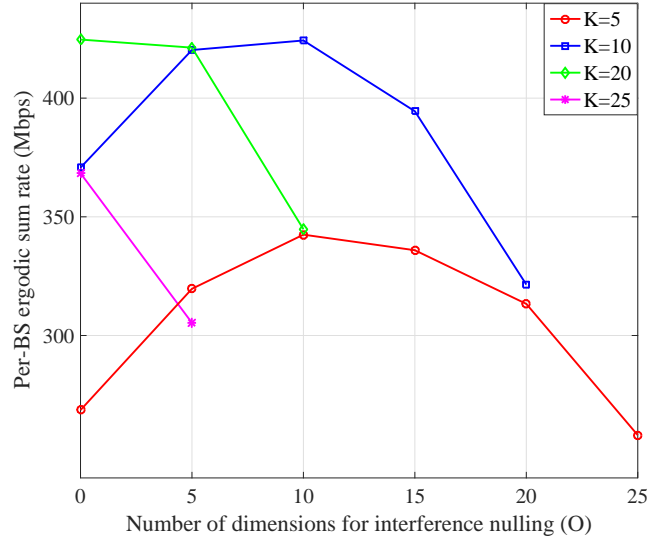}
\caption{Per-BS ergodic sum rate of an LS-MIMO system with $M = 32$, $K = 5$, $10$, $20$, $25$, and various choices $O$ obtained numerically. }
\label{fig:M32fixedSumRate}
\end{figure}

Fig.~\ref{fig:M32fixedSumRate} further supports the results in Figs.~\ref{fig:M15_optimum1} and~\ref{fig:M15_optimum2}. As in these figures, Fig.~\ref{fig:M32fixedSumRate} plots the per-BS ergodic sum rate for various values of $K$ as a function of $O$, but with $M=32$. Interestingly, even with a much larger number of available spatial degrees of freedom, amongst the cases tried, the highest sum rate is achieved with $\LB K^*,\zeta^*,O^*\RB = \LB 20,13,0\RB$. Again, as in the previous setting, with properly chosen parameters, out-of-cell interference nulling provides little benefit. 

It is important to emphasize that we are not suggesting that interference cancellation is never useful. As we will show in the next section, interference nulling can help cell-edge users.

\begin{figure}[t!]
\centering
\includegraphics[scale=1.05]{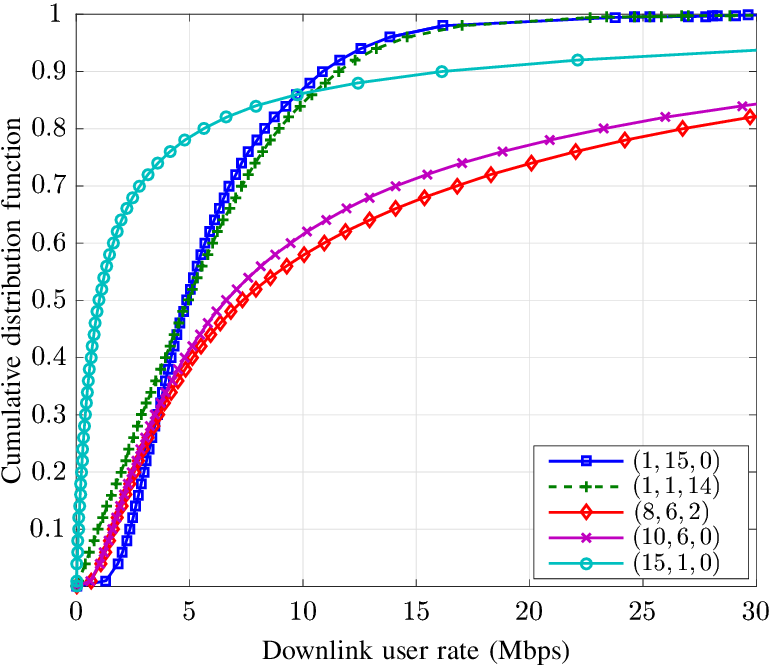}
\caption{CDF of downlink user rates for different choices of $\LB K,\zeta,O\RB$ with $M = 15$. }
\label{fig:per_cell2}
\end{figure}

Fig.~\ref{fig:per_cell2} plots the CDF of downlink user rates evaluated at various choices of $\LB K,\zeta,O\RB$, assuming a total of $K_b = 15$ users per cell. As can be seen, the gap between the CDF of the downlink user rates obtained at $\LB 10,6,0\RB$ and $\LB 8,6,2\RB$, the optimal triples from analysis and simulations respectively, is negligible. Fig.~\ref{fig:per_cell2} also plots the extreme scenarios of devoting all spatial dimensions to spatial multiplexing $\LB 15,1,0 \RB$, or to diversity $\LB 1,15,0 \RB$, or to interference nulling $\LB 1, 1, 14 \RB$. From a sum-rate perspective, our results have shown that operating at either $\LB 10,6,0\RB$ or $\LB 8,6,2\RB$ is superior to these extreme scenarios. However, interference nulling is seen to significantly benefit the cell-edge rate. In particular, if only one user is scheduled and all the BS antennas are used for nulling, i.e., for the $(1,1,14)$ scenario, significant cell-edge performance gain can be obtained. In the next section, we return to the question of how to optimize network parameters specifically for cell-edge users.

Table~\ref{table:optimal} lists the sum-rate optimal $\LB K^*,\zeta^*,O^*\RB$ obtained analytically as well as via numerical simulations for various values of $M$. The table also lists the performance gap between these operating points (evaluated by numerical simulations). In all the scenarios considered, the analysis indicates that it is optimal to ignore intercell interference and retain all spatial dimensions for multiplexing and diversity. The simulation results, in good agreement with the analysis, suggest that only very few of the spatial dimensions should be used for interference cancellation. Moreover, our results indicate that the performance gap between these points for various values of $M$ is minimal. 

\begin{table}[t!]
\caption{Optimal Operating Points $(K,\zeta,O)$ for Maximizing Per-BS Ergodic Sum Rate of LS-MIMO Systems}
\centering
    \begin{tabular}{|c|c|c|c|c|}
    \hline
    $M$ & Analytical & Numerical & Performance Gap & Loading Factor $\eta^*$\\  \hline
      $10$ & $\LB 6,5,0\RB$ & $\LB 5,5,1\RB$ & $4\%$ & $0.6$\\ \hline
    $15$ & $\LB 10,6,0\RB$ & $\LB 8,6,2\RB$ & $4\%$ & $0.66$\\ \hline
    $40$ & $\LB 25,16,0\RB$ & $\LB 24,15,2\RB$ & $3\%$ & $0.62$\\ \hline
         \end{tabular}
    \label{table:optimal}
\end{table}

An interesting observation from the table is that the optimal loading factor\footnote{The loading factor is defined as the ratio of the number of spatial dimensions used for both intra-cell and inter-cell interference cancellation to the total available number of spatial dimensions. The remaining number of spatial dimensions are used for providing diversity.} $\eta^* = \LB K^* + O^* \RB/M$ appears to be always close to $0.6$. Since $O = 0$ is approximately optimal, this implies that, for the parameters chosen and for round-robin scheduling, \emph{a close-to-optimum} operating point is for the BSs to devote about 60\% of its spatial resources to multiplexing ($K/M = 0.6$) and the remaining to providing diversity to each user ($\zeta = M-K+1$).

It is worth emphasizing that these results are obtained without accounting for the overhead imposed by channel training. Implementing interference nulling would increase the CSI acquisition overhead. As an example, with $M = 40$, at the optimal operating point obtained numerically, in order to serve $24$ scheduled users and null interference at $2$ out-of-cell users, $26$ channels must be estimated at each BS, whereas at the optimal point suggested by the analysis, i.e., without interference nulling, $K = 25$ channels must be estimated. Furthermore, with user-centric clustering, clusters may overlap. Therefore, the pilot allocation for CSI acquisition must be organized by some central entity across the entire network. This, in fact, represents another level of overhead. This suggests that in practical scenarios, enabling interference nulling may diminish, or even reverse, the apparent gain of $3$-$4\%$ by operating at the optimal point obtained numerically. However, it is equally important to note that due to the channel estimation uncertainty in practice, perfect CSI to the desired users as well as out-of-cell users is not available. The CSI estimation error not only affects the interference seen at out-of-cell users, but also causes intra-cell interference. The analysis of LS-MIMO systems under imperfect CSI is, however, beyond the scope of this paper.


\section{Interference Nulling and the Cell-Edge Rate}

The preceding section shows that, to maximize the \emph{sum rate}, it is close-to-best to spend the spatial dimensions for spatial multiplexing and diversity, and none for interference nulling. However, the CDF results also show that interference nulling does significantly help the users with low rate, e.g., users below the $10^{\mathrm{th}}$-percentile rate. While a user-centric clustering strategy negates the traditional concept of a ``cell", we refer to these low-rate users as ``cell-edge" users. In this section, we explore the optimal trade-off between multiplexing, diversity, and interference nulling for the metric maximizing the cell-edge rate, rather than the sum rate.  Cell-edge performance is a crucial consideration in system design, because from a wireless operator's perspective, the cell-edge rates relate to performance guarantees and, therefore, can be more important than the sum-rate performance.

\subsection{Range-Adaptive Clustering}

It should be noted that focusing on the cell-edge users leads to another potential variable to optimize. One of the assumptions made in the previous sections is that every user in the system chooses the same size of cooperating cluster (a constant $R_c$). This makes sense for sum-rate maximization, but if the objective is to maximize cell-edge rates, it may be beneficial to favor cell-edge users by increasing their cluster size. The reason is that the performance of the low-rate users is poor due to both weak signal power as well as strong inter-cell interference power. With a larger cluster size for cell-edge users, interference can be further mitigated and performance improved. Note that, to maintain the total number of dimensions available for interference nulling constant, we would have to decrease the cluster size of cell-center users. 

To facilitate the discussion, we call the original scheme of the previous section \emph{fixed-range interference nulling}, where all users have the same cluster size. We also devise \emph{range-adaptive clustering}, where different users have different cluster sizes. Ideally, the interference nulling range should be chosen according to the magnitude of the dominant interferers each user sees. As an approximation, this paper uses the heuristic scheme proposed in~\cite{li2015user} in which the cluster radius for each user is set to be proportional to its distance from its serving BS. The rest of this section compares the performance of fixed-range and range-adaptive clustering strategies.

Let $r$ denote the distance between a specific user and its serving BS; this distance has the CDF given in (\ref{F_r}). The cluster of the user covers a circle centered at the user. Let $R_c (r)$ denote the cluster radius of the typical user. In the range-adpative scheme, we set
\begin{equation}
R_c(r)=\nu r
\end{equation}
with an appropriate proportionality constant $\nu$.

To find the appropriate $\nu$, we use the fact that the total number of interference nulling requests sent by all users in an area must be equal to the total number of dimensions available for interference nulling at all BSs in the area. This relation, previously derived in (\ref{eq:average_B}), gives us that the average number of BSs in the cluster must be $\bar{B}=(O+K)/K$.

Now, let $B(r)$ be the number of BSs within the cluster radius of the typical user.  We know that since BS locations follow a PPP, $B(r)$ is a Poisson random variable with mean $\lambda\pi R_c(r)^2$ . By assumptions $A1$ and $A2$, to have $\mathbb{E}[B(r)]=\bar{B}$, we must have 
\begin{equation}
\int_{0}^{\infty} \lambda \pi (\nu r)^2 \mr{d} F_{r_{\min}}(r) = \frac{O+K}{K}.
\end{equation}
Hence, we arrive at
\begin{equation}
\nu = \sqrt{\frac{K+O}{K}}.
\end{equation}

\subsection{Interference Nulling for Cell-Edge Users}

We now examine the effect of different system parameters for cell-edge user performance via simulation. Unless specified, the parameters used are as listed in Table~\ref{table:parameters} and, as before, we fix $K_b = 15$ and use the scheduling fraction of $w = K/15$. 

\begin{figure}
\centering
\includegraphics[scale=1.05]{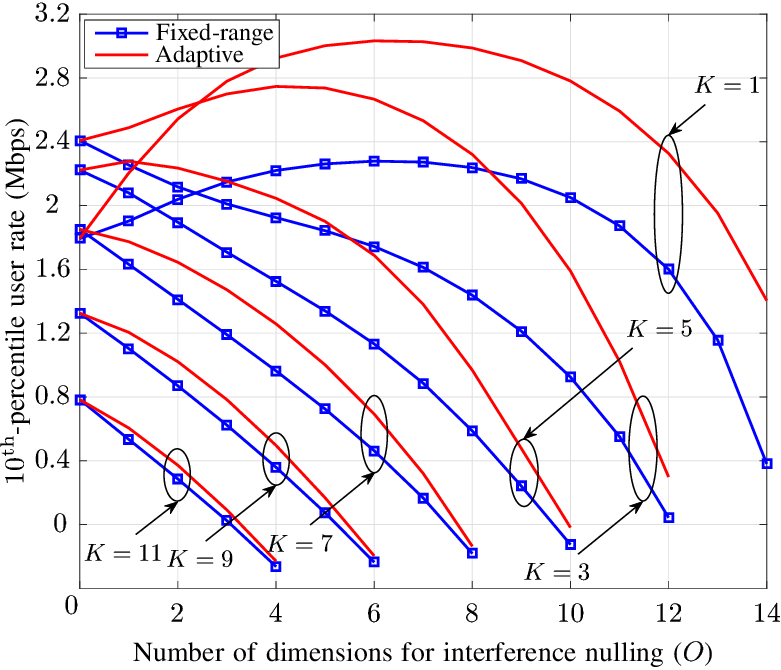}
\caption{The $10^{\mr{th}}$-percentile user rate comparison for LS-MIMO system
with $M = 15$, $K = 1, 3, 5, 7, 9, 11$, and various choices of $\zeta$ and $O$, under both fixed-range and range-adaptive clustering schemes.}
\label{fig:10_percentile_rate}
\end{figure}

Fig.~\ref{fig:10_percentile_rate} plots the $10^{\mr{th}}$-percentile user rate for both fixed-range and range-adaptive clustering schemes. The adaptive scheme clearly outperforms the fixed-range scheme for all network parameter choices. This is consistent with our intuition: range-adaptive clustering favors cell-edge users since these users are prone to significant interference; hence, they benefit from larger cluster sizes. We also see that the effect is especially pronounced when the number of users scheduled is small. In fact, the largest cell-edge rates are obtained with range-adaptive nulling with only one user scheduled per BS and about half of the dimensions used for diversity and half for interference nulling, i.e., the best operating point is $(1,8,7)$.

\begin{figure}[t!]
\centering
\includegraphics[scale=0.37]{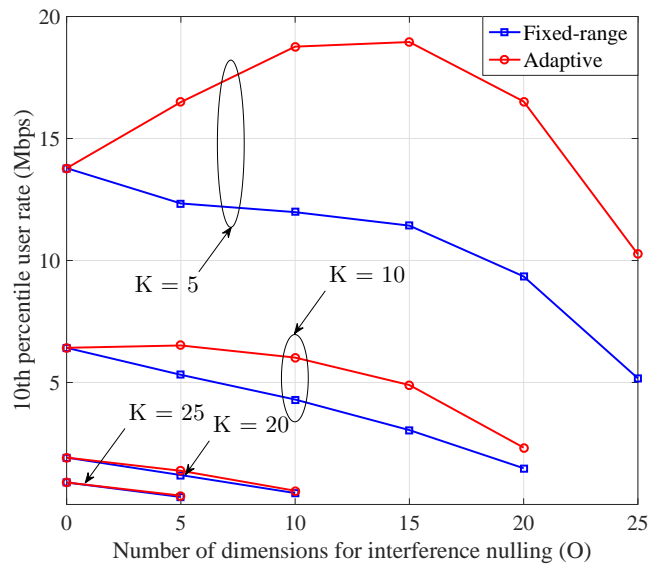}
\caption{The $10^{\mr{th}}$-percentile user rate comparison for LS-MIMO system
with $M = 32$, $K = 5, 10, 20, 25$, and various choices of $O$, under both fixed-range and range-adaptive clustering schemes.}
\label{fig:M3210_percentile_rate}
\end{figure}

Fig.~\ref{fig:M3210_percentile_rate} provides a similar result as the previous figure for the case of $M=32$. As before, range-adaptive nulling significantly improves the 10th-percentile rate, again, for small values of $K$, the number of users served. The largest rate is obtained with $K=5$ users served within the cell while interference is nulled at $O=15$ out-of-cell users. 

\begin{figure}
\centering
\includegraphics[scale=1.05]{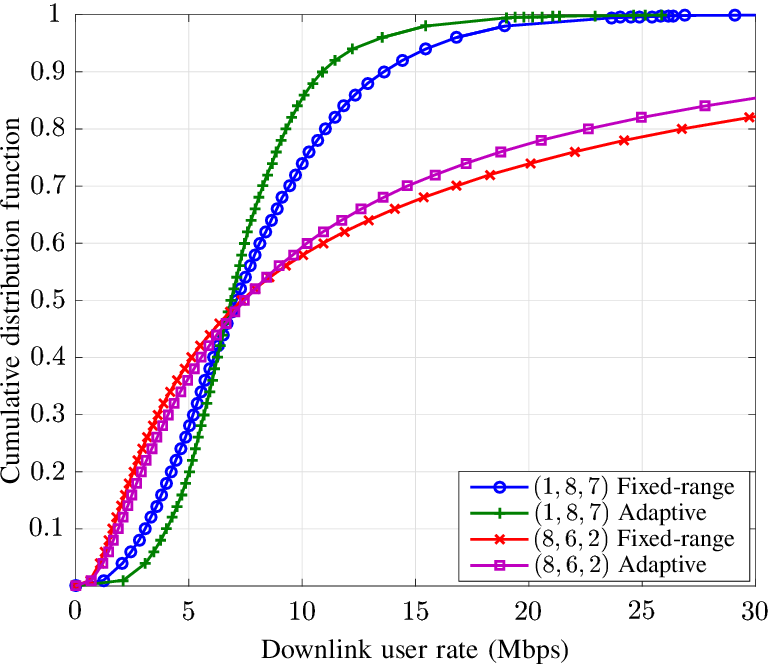}
\caption{CDFs of downlink user rates for the sum-rate optimal $(K,
\zeta, O) = (8,6,2)$ and the cell-edge rate optimal $(1,8,7)$ schemes
under fixed-range and range-adaptive clustering.}
\label{fig:CDF_constant_dynamic}
\centering
\vspace*{0.2in}
\includegraphics[scale=1.05]{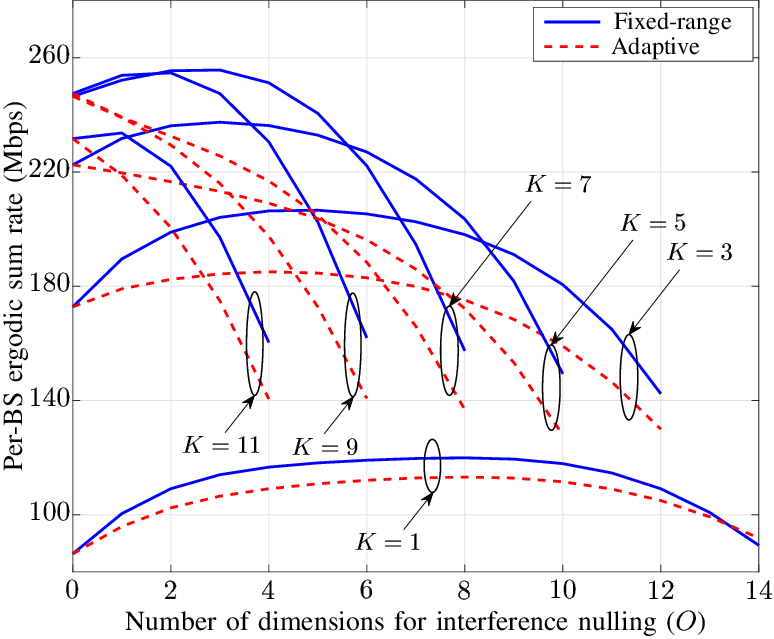}
\caption{Per-BS ergodic sum rate of an LS-MIMO system versus $O$ with
$M = 15$ and various choices of $K$
under both fixed-range and range-adaptive BS clustering schemes.}
\label{fig:ergodic_rate_constan_dynamic}
\end{figure}

To explicate the effect of range-adaptive clustering for all users, Fig.~\ref{fig:CDF_constant_dynamic} plots the CDF of user rates for both the sum-rate optimal $(8,6,2)$ scheme as well as the edge-rate optimal $(1,8,7)$ scheme under both the fixed-range and range-adaptive clustering schemes. The figure shows that the range-adaptive scheme benefits low rate users at the expense of high rate users. This is expected because the total number of dimensions available for interference nulling is fixed---the larger number of interference nulling requests by cell-edge users means that fewer requests by cell-center users can be accommodated. The range-adaptive scheme in fact reduces the effective cluster size for cell-center users, thus degrading their performance, while improving the rates for the cell-edge users.

Does range-adaptive clustering improve the sum rate? To answer this question, we evaluate the per-BS ergodic sum rate performance in Fig.~\ref{fig:ergodic_rate_constan_dynamic}. This figure shows that range-adaptive clustering reduces the sum-rate as compared to fixed-range interference nulling. This suggests that the benefit of range-adaptive nulling for cell-edge users does not outweigh its cost to the cell-center users, thereby reducing the sum rate. It is interesting to observe from Fig.~\ref{fig:ergodic_rate_constan_dynamic} that for range-adaptive nulling, it is in fact optimal to use all the spatial dimensions for multiplexing or diversity, and none for interference nulling, in order to maximize the sum rate. Similar to the fixed-range case, the optimal loading factor remains about 60\%.

\subsection{The Impact of User Distribution}
\label{subsec:normalized}

Our final result addresses one open issue in our development. Based on Section~\ref{sec:simul}, in the analysis and the numerical results so far, we have assumed that every user throughout the network is scheduled for the same fraction of time. In practice, however, this fraction would be related to the overall number of users within the cell. Specifically, the fraction available to each user in cell $b$ is $K/K_b$ where $K_b$ is the \emph{total number} of users in that cell. In this final result, we show that accounting for this variation across the cells does not change our conclusions. Since the sum rate is not affected by this fraction, we focus on the user rate CDF. 

\begin{figure}
\centering
\vspace*{0.2in}
\includegraphics[scale=1.05]{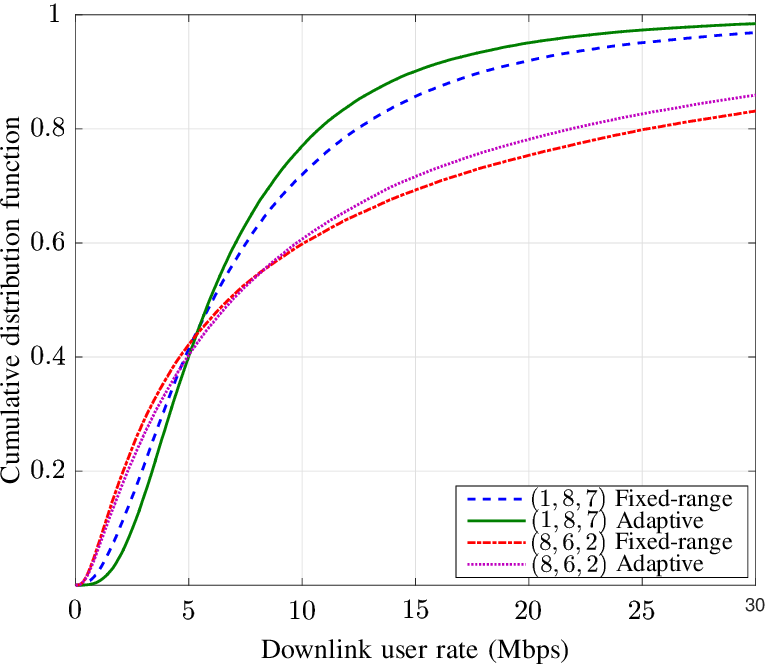}
\caption{CDFs of downlink user rates for the sum-rate optimal $(K,\eta,O) = (8,6,2)$ and the cell-edge rate optimal $(1,8,7)$ schemes under fixed-range and range-adaptive clustering, while accounting for the different total number of users in cells of different sizes.}
\label{fig:normalized}
\end{figure}

Fig.~\ref{fig:normalized} plots the user-rate CDFs, obtained via simulation, for two choices of the triple $(K,\zeta,O)$: the triples that provide the maximum per-BS ergodic sum rate, (8,6,2), and the maximum cell-edge rate, (1,8,7). These are the same cases as in Fig.~\ref{fig:CDF_constant_dynamic}; the only difference is that, here, the scheduling fraction is different in each cell and set to be $K/K_b$ where $K_b$ is the realization of the number of users in the cell. To make the comparison fair, we fix the same average user density as before. The plot presents the CDFs for both the fixed and range-adaptive clustering strategies. Comparing Figs.~\ref{fig:CDF_constant_dynamic} and~\ref{fig:normalized}, it is clear that the comparison between the CDFs remains essentially the same as before. The theory and discussions of the previous sections are, therefore, valid despite the use of a fixed scheduling fraction. 

\section{Conclusions} \label{sec:summary_chap4}

Multiplexing, diversity, or interference nulling? When BSs in a network are equipped with a large number of antennas, the choice of how to utilize these spatial dimensions has a significant effect on the overall system performance.  This paper considers what we call an LS-MIMO system, a system with sufficient antennas so all three uses can occur simultaneously; the system uses user-centric clustering operating under ZF beamforming and equal power assignment. We show that the answer to this question depends on the system objective being optimized.

When the system objective is to maximize the ergodic sum rate, the analysis in this paper shows that at the optimal operating point, only very few of spatial dimensions should be reserved for interference nulling. It is, in fact, near optimal to allocate none of the spatial resources for interference nulling, while using a fraction of spatial dimensions for multiplexing (assuming that there are a sufficiently large pool of users) and the rest for providing diversity. Our simulations show that reducing multiplexing and diversity dimensions in order to perform interference nulling does not appear to bring much benefit to the overall system sum rate.

This conclusion is supported by both our analytical results based on stochastic geometry, which provide computationally efficient expressions for the per-BS ergodic sum rate as a function of average cluster size, as well as system-level simulations. The analytic results allow us to optimize the per-BS ergodic sum rate of the system as a function of the number of users to schedule, the diversity order of each user and the number of interference directions to cancel.

When the system objective is to maximize the cell-edge rate, 
this paper shows that it is optimal to schedule just one user per cell, and use about a half of the remaining dimensions for diversity and the other half for interference nulling. In addition, a range-adaptive nulling strategy can outperform fixed-range interference nulling. However, such an interference nulling strategy comes at significant cost to the achievable per-BS ergodic sum rate in an LS-MIMO system.


\bibliographystyle{IEEEtran}
\bibliography{IEEEabrv,nulling}

\end{document}